\documentclass[manuscript,screen,nonacm]{acmart}
\setcopyright{none}

\usepackage{xspace} 
\usepackage{tabularx}
\usepackage{xcolor}
\usepackage{color}
\usepackage{framed}
\usepackage{fancybox}

\usepackage{bbding}

\newenvironment{coloredleftbar}[1]{%
  \def\FrameCommand{\colorbox{#1}}%
  \MakeFramed {\advance\hsize-\width \FrameRestore}}%
{\endMakeFramed}

\AtBeginDocument{%
  }

\acmDOI{XXXXXXX.XXXXXXX}

\begin{document}

\title[Design Requirements for Patient-Centered Digital Health Applications]{Design Requirements for Patient-Centered Digital Health Applications: Supporting Patients' Values in Postoperative Delirium Prevention}


\author{David Leimstädtner}
\email{david.leimstaedtner@fu-berlin.com}
\orcid{0000-0002-1445-3153}
\affiliation{%
\institution{Freie Universität Berlin}
\city{Berlin}
\country{Germany}
}

\author{Fatima Halzl-Yürek}
\email{fatima.yuerek@charite.de}
\orcid{0000-0002-3739-4619}
\affiliation{%
\institution{Charité – Universitätsmedizin Berlin}
\city{Berlin}
\country{Germany}}

\author{Claudia Spies}
\email{claudia.spies@charite.de}
\orcid{0000-0002-1062-0495}
\affiliation{%
\institution{Charité – Universitätsmedizin Berlin}
\city{Berlin}
\country{Germany}}

\author{Claudia Müller-Birn}
\email{clmb@inf.fu-berlin.de}
\orcid{0000-0002-5143-1770}
\affiliation{%
\institution{Freie Universität Berlin}
\city{Berlin}
\country{Germany}}

\renewcommand{\shortauthors}{Leimstädtner et al.}

\begin{abstract}
Postoperative delirium (POD) is among the most common complications after surgeries for older adults and can entail long-term adverse health consequences. Active patient participation in POD prevention presents a central factor in reducing these risks.
To support patient engagement through a digital health application, we use value sensitive design approaches to identify the requirements for a patient-centered digital health application supporting patient engagement in POD prevention.
Through interviews with medical professionals and patient representatives, we construct a patient journey, which serves as the basis for twelve patient value journey interviews. In these interviews, patients from the high-risk group for POD revisit their recent experience of undergoing surgery to elicit barriers, needs, and values concerning POD prevention from a patient perspective.
An analysis of the patient interviews derives four design requirements for a digital health application supporting patients regarding POD prevention: the adaptation of patient-centered communication, the provision of procedural transparency, fostering patient empowerment through consistent guidance, and explicitly addressing relatives as mediators and supporters for a patient after a POD occurrence.
\end{abstract}



\received{XX XX 2025}

\maketitle
\section{Introduction}
Postoperative delirium (POD) is among the most commonly occurring complications after surgeries for older adults~\cite{yurek_qualitatsvertrag_2023, milisen_is_2020}.
It describes a state of acute confusion characterized by attention deficit, impaired consciousness, and thought disorders~\cite{swarbrick_evidence-based_2022}.
In the long term, POD is linked to adverse health outcomes for affected patients, including cognitive and functional decline, higher mortality rates, and an increased risk for institutionalization in care facilities~\cite{milisen_is_2020}.
POD may occur after a surgery~\cite{yurek_qualitatsvertrag_2023}, yet remains reversible if proper interventions and treatments are applied~\cite{milisen_is_2020, swarbrick_evidence-based_2022}.
Delayed initiation of such treatments, however, is associated with increased mortality, a higher likelihood of needing post-hospital care, and the development of neurocognitive disorders~\cite{yurek_qualitatsvertrag_2023}.
Considering the seriousness of these consequences, POD prevention measures\footnote{In the context of this paper, \textit{measure} refers to a course of action taken to achieve a particular purpose (e.g., POD prevention).} serve to safeguard a patients's health, cognitive, functional, and mental functions and support their autonomy~\cite{yurek_qualitatsvertrag_2023}.
However, despite evidence highlighting the efficiency of preventive measures in reducing the occurrence of POD and severity of related long-term effects, these measures have long not been integrated into the clinical care routine~\cite{swarbrick_evidence-based_2022}.

Addressing this gap, European guidelines on evidence-based care for POD have been updated in 2024 to ensure adequate prevention and treatment strategies~\cite{aldecoa_update_2024}.
In medicine, a significant concern is investigating how a digital clinical decision support system can integrate these guideline recommendations into the clinical routine. A central challenge in developing such systems is integrating at-risk patients as active stakeholders in POD prevention activities.
Active integration of patients presents a crucial component of POD-preventive strategies~\cite{yurek_qualitatsvertrag_2023}.
Patient-centered design in healthcare can enhance patient engagement and empowerment, instigating surgical patients to assume an active role in their journey to well-being during and beyond their hospitalization, rather than passive care recipients. Digital health systems, for example, patient portals or interactive educational resources, can instill such a sense of agency for patients~\cite{langote_humancomputer_2024}.
Yet, such patient-sided digital health applications also entail an expansion of healthcare work from a relatively bounded clinical setting into an overarching effort spanning organizational and professional contexts.
In the case of POD, this entanglement of clinical guidelines, collaborative care, and patient's social and technological circumstances presents a \textit{complex intervention}, where a computer-supported cooperative work (CSCW) perspective can serve to evaluate this emerging healthcare technology (i.e., decision supports system for POD guidelines) with respect to \textit{ ``the ways in which processes are dynamically situated in, and contingent on, local practices and contexts''}~\cite{fitzpatrick_review_2013}.
In this regard, improving patient engagement and communication with digital healthcare processes has been an active research area in CSCW~\cite{berry_supporting_2021}.
In designing supportive measures of patient engagement in emerging digital health technologies, it is crucial to recognize how the vulnerability of this stakeholder group may lead to an underrepresentation of their values in new health technologies.
Digital health technology is not neutral but rather embeds the values of its stakeholders to differing degrees~\cite{tseng_data_2024,cajander_electronic_2019,sorries_advocating_2024}. The values of vulnerable stakeholders, such as patients, may be overlooked due to power differentials in the design processes.
Hence, to ensure that patients' needs and values are met when designing digital health technology, involving these stakeholder groups is crucial~\cite{grunloh_and_2024}.
Adapting value sensitive design (VSD)~\cite{friedman_value_2019} (discussed in detail in \autoref{sec:vsd}) provides a methodological framework for recognizing these stakeholder values and utilizing them as the basis for the design of digital health technology~\cite{sorries_advocating_2024}.
%

Against this background, we present our approach to designing a patient-centered digital health application that supports POD prevention.
Therein, the guiding research questions are as follows:

\begin{description}
    \item[\emph{RQ 1:}] \emph{What are patients' values and barriers concerning POD and related prevention measures when undergoing surgery?}
    \item[\emph{RQ 2:}] \emph{What are the design requirements for a digital health application that supports patients' values in overcoming barriers toward engagement in POD-preventive measures?}
    
\end{description}

To address the research questions, we integrated previous work from the fields CSCW and human-computer interaction (HCI), to approach digital health technology design from a patient-centered perspective, referring to VSD~\cite{friedman_value_2019} framework and the related \textit{patient value journey maps}~\cite{bui_patient_2023,grunloh_and_2024} approach, to account for patient values in designing supportive digital health technology in the application context POD prevention.
Based on this theoretical and methodological background, we first conducted stakeholder interviews with medical professionals and patient representatives to outline the current practices and processes related to POD-prevention measures during a surgical patient's hospitalization. 
Second, patient value journey interviews were conducted to investigate their experiences.
A thematic analysis identified patient barriers and values concerning POD that are consequently operationalized into design requirements for a future patient-centered digital health application. 
%
%
The derived design requirements constitute what \citet{hook_strong_2012} describes as \textit{`intermediate level design knowledge'}, articulated through \textit{`strong concepts'}. 
Strong concepts present core design ideas for interactive behavior that are abstracted beyond specific design instances yet below generalizable theory. As such, they are generative pieces of knowledge, rooted within a deep understanding of the particulars of a specific design situation, providing direction in the generation of new solutions therein~\cite{hook_strong_2012}.
Such an approach corresponds to what \citet{schmidt_permutations_2007} argues for CSCW researchers to support both local practices and develop higher-order practices in the sense of standardized building blocks suitable for more generalizable reuse in future research efforts (also see discussion in \citet{fitzpatrick_review_2013}). 
Hence, our research offers the following contributions to the CSCW community:
\begin{enumerate}
\item \textit{\textbf{Contextualized Understanding of Patient Barriers and Values on POD}}
Through interviews with different stakeholders within the research context, we provide a deep, contextualized understanding of patients' values concerning POD. 
For this purpose, we first establish the typical patient journey within the application context. 
Second, we elicit values and barriers along this journey through \textit{patient value journey interviews} with patients from the high-risk group, reviewing their recent experience of undergoing surgery.

\item \textit{\textbf{Value-Centered Method to Deriving Design Requirements for Patient-Centered Digital Health Applications}}
Through a thematic analysis of the patient value-journey interviews, we operationalize the barriers and values derived from patients' experience into design requirements for supporting patient engagement in POD prevention through a digital health application. 
These \textit{strong concepts} provide guidance for the future design of digital health technologies, supporting POD preventive measures based on patient values.
We position these findings regarding patient-centered technology and patient values within ongoing discourse in CSCW, and suggest future directions for supporting patient values in digital health.
\end{enumerate}

%

\section{Related Work}
\label{sec:relatedwork}
In the following, we (1) outline considerations from CSCW and HCI toward designing patient-centered digital health technology, (2) highlight VSD as a methodological means to realize stakeholder values in design, and (3) review the challenges toward supporting patient engagement concerning POD when undergoing surgery.

\subsection{Designing Patient-Centered Digital Health}
In medicine, the concept of \textit{patient-centered care} looks at providing individualized support considering a patient's personal needs, with previous research demonstrating the capability of such a healthcare approach in enhancing patient engagement and health outcomes~\cite{jacobs_designing_2015}.
The challenges to consistently adopting a patient-centered approach lie in healthcare providers' constraints regarding the time and resources needed to provide such individualized patient support. 
\citet{jacobs_mypath_2018} highlighted this gap as a valuable opportunity for efforts in the development of patient-focused digital health tools, fostering individualized support for patients.
Concerning the design of supportive digital health technology, following such a patient-centered perspective means designing with patients' physiological and psychological perspectives in mind~\cite{gao_information_2019}.
Also within CSCW, research concerning digital health has seen a shift in focus concerning the the role of the patient there-in: \citet{fitzpatrick_review_2013} described this move from the patient a passive actor, that is mostly invisible apart from being the object of the information and workplace coordination efforts toward the patient being a key stakeholder, taking an active role through engagement in healthcare decision-making and treatment. Hence, recent research efforts have investigated how digital health applications can support patients therein: 
\citet{park_patient_2017} emphasized how mismatches between health literacy and knowledge of healthcare practices between patients and providers create conflicting perceptions, rendering patients unable to access, interpret, and use relevant information required for adopting an active role. The authors investigated the use of patient-side information technologies, such as wall displays, tables, kiosks, or mobile apps, as a means to provide them with care-related information, enhance patient engagement, and improve communication with healthcare providers.
%
%
In HCI, \citet{cajander_electronic_2019} highlighted how digitalization in healthcare enables patient participation: They found that integrating patients into processes, for example, by allowing them to access their health information via a digital application, increased patient involvement and overall quality in care. Furthermore, this involvement also motivated patients to actively contribute to their healthcare processes by adding information they deemed relevant to their health records.
A meta-analysis by \citet{sawesi_impact_2016} confirmed this role of IT platforms in enhancing patient engagement. Further, the authors found that such active patient participation was correlated with greater improvements in health status outcomes compared to less involved patients.
Yet, as outlined in \citet{fitzpatrick_review_2013}'s review of the past 25 years of research on healthcare in CSCW, the complexities inherent to new models of care brought about by digitalization of the healthcare system require researchers to engage with this tension in the collaboration and information sharing between localized practices (i.e., at a particular clinic) and standardized means (e.g., patient-portals available to a broad public).
While health technologies can support individuals' health behavior, usage patterns and requirements vary significantly between user groups (e.g., older adults). The design of digital health applications needs to consider user capabilities. 
In this regard, previous work has leveraged qualitative inquiries to highlight the perspective and values of older users on particular medical conditions~\cite{harrington_informing_2018} for embedding health technology design with a patient perspective.

In summary, while advancements in digital health hold great potential for increasing patient engagement and, consequently, improving health outcomes, considering patients' needs is crucial to ensuring that these benefits can be realized. Especially in a context like POD, where the high-risk group consists of adults over 75 years old, the design of a digital health application must be based on the specific needs and values of this target group.

\subsection{Value Sensitive Design and Patient Values}
\label{sec:vsd}
Value sensitive design (VSD)~\cite{friedman_value_2019} is a theoretically grounded approach to technology design that focuses on the values of those who actively engage with technology or are affected by it. 
Values are thereby understood as \emph{``what is important to people in their lives, with a focus on ethics and morality~\cite{friedman_survey_2017}}.
By investigating human values, VSD aims to inform technology design to be representative of its stakeholders' values.
Previous work in VSD has characterized values as situated and contextualized, shaped by the lived experience of stakeholders~\cite{so_they_2024}.
In work derived from VSD, \citet{jonas_designing_2022} laid out that researchers must consider potential conflicts between their values and those of a vulnerable participant group. To achieve the representation of such a stakeholder group's values, design efforts need to be embedded within a deep understanding of the research context.
In this regard, \citet{le_dantec_values_2009} argued that it matters how researchers approach values, as an initial focus on preconceived, theoretically derived values can potentially blind the researcher to values that fall outside. 
Comparably, \citet{iversen_rekindling_2010} proposes an emergent approach to unearth contextualized values for technology design.
Engaging in such initial empirical investigation to discover stakeholder values can serve as a starting point for value-sensitive design efforts, helping to avert the researcher bias~\cite{le_dantec_values_2009}.
\citet{sorries_advocating_2024} conceptualized value-elicitation activities for patients that serve to emerge contextualized values toward the design of value-centered digital health applications by supporting patients in reflecting on their values for a given design context in digital health. Here, value elicitation is achieved through a value questionnaire, prompting patients to consider the role of a particular value in a specific medical application context.
\citet{bui_patient_2023, grunloh_and_2024} proposed a \textit{patient journey value mapping}, an approach to capturing experiences, emotions, and values implicated in patients' care delivery experience. This approach acknowledges that patients' values may change across different phases of their journey and that designers should respond to these evolving needs when designing digital health technology. To achieve this, the authors employed a paper-based mapping concept that tracks patients' values and their respective prioritization over time, allowing designers a deeper understanding of patients’ experiences and values to identify unmet patient needs and design opportunities for patient-centered improvements in digital health. 

A core theoretical supposition of VSD is the interactional character of the relationship between technology design and social context, wherein the design features of a given technology always either support or undermine particular stakeholder values. This interplay between users and context ultimately determines the influence of technology design on their larger application context (or society at large)~\cite{riebe_values_2024, davis_value_2015}.
Previous work has established that the substantial power differentials and information asymmetries inherent to patients' roles in healthcare can have an impact on technology design in digital healthcare~\cite{cajander_electronic_2019}.
%
%
Values related to the patient's needs, their philosophy of life, and background are often disregarded, as medical practitioners tend to prioritize values directly associated with patient treatment~\citet{lee_exploring_2013}. 
Patient-centered and value-sensitive approaches can make these underlying values visible, enabling their integration into the design of future technologies that support patient engagement in digital health processes. This is especially important for vulnerable groups such as older adults at risk for POD, where accommodating this specific group's needs is crucial~\cite{anuyah_characterizing_2023, harrington_informing_2018}
Designing equitable health technologies that are tailored to the unique needs of older adults is critically important, as they often struggle with digital health tools that overlook their values and preferences~\cite{so_they_2024}. A misalignment between users' values and those embedded in the technology can result in rejection by the intended user group~\cite{scheuerman_datasets_2021}.
\citet{anuyah_characterizing_2023}'s synthesis of HCI work with vulnerable target groups echoed a call from these research communities to consider engagement methods that can carefully empower vulnerable groups. The underscored the use of frameworks like VSD as a means to guide engagement with vulnerable populations without overburdening the participant groups~\cite{anuyah_characterizing_2023}.
In this regard, recent work in HCI on designing for older adults has seen a shift in focus from deficiency due to barriers and accessibility challenges toward a user-centered approach, aiming to take older adults' wants, needs, desires, and expectations into account as the underlying basis for design~\cite{hao_exploratory_2023}. Researchers have employed diverse methods, including surveys, interviews, and focus groups, to gather older adults' needs and design considerations, thereby fostering a proactive approach emphasizing empowerment by involving older adults in the technology development process~\cite{zhao_older_2024}.

In summary, VSD offers a design approach for embedding stakeholder values into technology design. For patient-centered digital health, VSD allows the consideration of patients' unique needs and contexts, addressing power imbalances and value misalignments. Specifically, Patient Value Journey Mapping helps consider the evolving values throughout the surgical patient journey.

\subsection{Supporting Patient Engagement for POD}
Patient engagement refers to the active collaboration between patients and healthcare professionals to improve patient care~\cite{haldar_beyond_2019}. 
\citet{haldar_beyond_2019} attributed several positive outcomes to increased patient engagement, including improved clinical outcomes, increased understanding of their health condition, reduced risk for medical errors, and strengthened patient-provider relationships. 
This extends to POD prevention strategies, where active patient engagement presents a central component~\cite{yurek_qualitatsvertrag_2023, swarbrick_evidence-based_2022}.
Health information technologies can play a crucial role in promoting patient engagement, as demonstrated across various patient populations (see ~\citet{haldar_beyond_2019}). 
\citet{langote_humancomputer_2024} further highlighted how interactive healthcare interfaces empower patients to engage as active agents of their own healthcare rather than passive recipients.
Patients' autonomous and self-guided engagement with their health situation, for example,  by accessing health data during ongoing healthcare, has been associated with better health outcomes and patient satisfaction~\cite{park_beyond_2017}, including POD~\cite{yurek_qualitatsvertrag_2023}.
This is further illustrated in \citet{pfeifer_vardoulakis_using_2012}'s study, which provided hospitalized patients with insights into their care via a mobile application and found that patients experienced an increased sense of participation in their care. The authors suggest that this approach holds great promise for reducing patient anxiety, raising awareness, and increasing patient empowerment, along with patients' perceptions of ownership of their care.
Yet, in the clinical context, patients commonly experience barriers towards active engagement in POD measures:
%
%
Although patients are routinely provided with verbal and written information about POD during preoperative consultations, the information provided in such meetings is frequently not adequately perceived by patients. This is demonstrated across several studies, which found that patients' recall of medical information after a hospital visit is low~\cite{wilcox_designing_2010}.
The reasons for this are manifold and can include, for example, the application of terminology and information modes that remain inaccessible to the patient, resulting in mismatching expectations between the parties~\cite{wilcox_designing_2010}, or the influence of asymmetrical power relations, which may hold patients from gaining suitable levels of information~\cite{dahl_facilitating_2020}.
This mismatch of the patient-provider communication is also identified by \citet{park_patient_2017}, who describes this mismatch as inhibiting patients from taking an active role.
Concerning hospitalized patients, \citet{haldar_beyond_2019}'s interview study, including both patients and caregivers, found that patients perceive the information they received on care decisions to be lacking. They expressed their information needs regarding future procedures, the reasoning behind these, associated risks and benefits, and how long they would take. Further, patients reported challenges in asking questions, as they were unsure about who to ask within a care team and did not want to overburden its members~\cite{haldar_beyond_2019}.
Taken together, these issues underscore why implementing measures designed to facilitate patient engagement can serve a key role in supporting POD prevention measures.

%
\textit{So, what patient needs ought to be facilitated to activate patient engagement in POD preventive activities?}
Focusing on the patient experience when faced with the prospect of undergoing surgery, \citet{suhonen_adult_2006} emphasizes how the availability of adequate information is necessary to allow patients to engage in appropriate care management and coping, as well as for reducing stress and anxiety associated with illness and surgery.
Recent work on POD prevention found that satisfying patients' information needs serves to reduce POD-related anxiety~\cite{chen_design_2017}.
Education programs for patients and their families, aimed at increasing pre-existing knowledge about a particular condition, can improve patient experiences~\cite{chen_design_2017}. Substantially, these education efforts concern surgical procedures, possible outcomes, and preemptive exposure to the surgical environment they will experience~\cite{swarbrick_evidence-based_2022}.
\citet{xue_preoperative_2020}, for example, confirmed the effect of individualized preoperative information approaches, finding that patients undergoing cardiac surgery have a lower incidence of POD for individuals after receiving individualized education sessions, a critical care tour, as well as an information leaflet compared to those receiving standard care. 
Such educational measures have also been proposed as facets of multi-component interventions for POD prevention. In this regard, an expert consensus paper by \citet{peden_improving_2021} endorses offering patients and their relatives education about POD-associated risks, prevention-supporting techniques, and information about possible long-term health effects as a measure to reduce delirium incidence~\cite{swarbrick_evidence-based_2022, peden_improving_2021}.

In summary we outlined how, prior research across CSCW, HCI, and digital healthcare has indicated that effective communication between patients and their providers increased both patient satisfaction and health outcomes~\cite{park_patient_2017}, highlighting how patients being engaged in and informed about their care has a positive impact on clinical outcomes~\cite{pfeifer_vardoulakis_using_2012,yurek_qualitatsvertrag_2023}. Yet, patients are commonly insufficiently informed, even about fundamental aspects of their care~\cite{pfeifer_vardoulakis_using_2012}.
Previous findings indicate that while patients facing surgery would need information to engage with POD proactively, the clinical realities of communicating this information to the patient can often make this difficult. Therefore, patients should be supported in their specific needs (e.g., concerning information provision) to better promote patient engagement.

\section{Research Context}
\label{sec:context}
This research effort is part of a larger research consortium developing a digital decision support system that makes current evidence-based guideline recommendations for POD machine-readable and enables automated, real-time validation of treatment actions taken by clinical staff. 
The project aims to reduce barriers to implementing current guideline recommendations, decrease clinic workload, enhance efficiency in treatment processes, and close care gaps, thereby improving patient safety. 
This system is being developed and evaluated at one of Europe's leading academic medical centers, which plays a pivotal role in advancing digital health strategies to address POD among older surgical patients. Germany's aging population and increased risk for POD incidences underscore the importance of this research endeavor~\cite{yurek_qualitatsvertrag_2023}.
The authors of this paper were responsible for designing a user interface that provides guideline-based information to patients and supports their participation in POD prevention measures.
To ensure that patients' needs and values are met when designing digital health technology, involving these stakeholder groups is crucial~\cite{grunloh_and_2024}. 
Yet the at-risk group for POD (i.e., surgical patients above the age of 70) poses a particularly vulnerable target group, requiring utmost methodological consideration to devise suitable approaches toward embedding their perspective into the design of future technology.
\citet{boulus-rodje_ageing_2015} highlighted the need for design research working with older adults to employ contextualized, empirical research efforts to approach this group's heterogeneous and diverse nature in terms of experiences, needs, and capabilities to derive healthcare technologies that are adaptable and relevant to different individuals
In consideration of the situated nature of digital health applications, qualitative and contextualized research approaches offer a means to ground patient-centered design of technology in the sociotechnical reality in which end users will engage with it~\cite{kassam_patient_2023, tseng_data_2024}.

\subsection{Positionality Statement}
\label{sec:positionality}
We recognize the centrality of the researcher's interpreting role in qualitative, contextualized, and VSD. Therefore, we follow \citet{borning_next_2012}'s recommendation for explicating the involved authors' backgrounds by describing their positionality.
The author team comprises two HCI researchers from a German university's computer science department and two medical professionals at the collaborating university hospital, which specializes in POD.
Authors 1 and 4 are part of a research group that pursues a human-centered approach to investigating the underlying assumptions and implications of technology design in digital health. The group focuses on digital healthcare and supports vulnerable stakeholder groups in technology design. Author 1 has a background in communication science and Author 4 a background in computer science, both focus on human-computer interaction in digital health. Both authors are white and Western European. We acknowledge that these cultural and socioeconomic backgrounds may influence our understanding and framing of the issues studied in this article. Neither author has previous experience concerning POD.
As part of the larger research project (described above in \autoref{sec:context}, this study involves collaboration with medical researchers:
Author 2 is a senior physician and an anesthesiology specialist. Author 3 is the director of an anesthesiology and intensive care medicine clinic. Their role in this study included supporting gaining ethics approval from the hospital's ethics review board, assigning study nurses to support the recruitment of interview respondents, consulting about the medical factors described concerning POD measures, and reflecting with Authors 1 and 2 on their findings from a medical and university hospital perspective.

\section{Method}
\label{sec:method}
To gain a comprehensive understanding of the POD-prevention application context and to identify resulting design considerations for a digital health application supporting patients toward engagement in POD-prevention activities~\cite{bodker_what_2022}, this study employs a qualitative design using interviews with multiple stakeholder groups.
Specifically, we conducted (1) semi-structured interviews with medical professionals and patient representatives (referred to as \textit{background interviews}), followed by (2) patient value journey interviews with patients who had undergone surgery in a clinic with POD screening measures within the preceding year.
%
%
Ethical approval for this study was obtained from the ethics committee of the collaborating university hospital~\textit{(Name removed for anonymization purposes)}.

\subsection{Participants and Recruitment Process}
In the following, we detail recruitment processes for both interviews. An overview of all respondents is provided in \autoref{tab:all_participants}.

    \begin{table}[ht]
    \caption{Overview of all participant groups across interview types.}
    \centering
    \begin{tabular}{lll}
    \toprule
    \textbf{Respondent Group} & \textbf{N} & \textbf{Interview type }\\
    \midrule
    Patients & 12 & Patient Value Journey Interview \\
    Patient Representatives & 3 & Semi-Structured Background Interview\\
    Anesthesiologists & 4  & " \\
    Nursing staff & 3  & " \\
    Delirium experts & 1  & " \\
    \bottomrule
    \end{tabular}
    \label{tab:all_participants}
    \end{table}

\paragraph{Clinical Staff and Patient Representatives}
\label{sec:recruitment_background_interviews}
Medical staff and patient representatives were interviewed to gain a comprehensive picture of the patient journey regarding POD. 
The interview partners were recruited through purposive sampling within the application context (i.e., the research consortium). 
A range of healthcare professionals at the participating university hospitals were identified as being involved in various roles in POD education, prevention, monitoring, treatment, and aftercare activities. These include four anesthesiologists, three nursing staff members, and one delirium expert. 
The patient representatives recruited for this purpose included a government official designated as the patient representative for the city government of a large German city, an associate from a local network for patient safety, and an engaged relative caring for their husband, who lives in a care facility after a POD occurrence.
Participants did not receive monetary compensation for participation.
%
%
Participants did not receive monetary compensation for participation.

\paragraph{Patients}
\label{sec:recruitment_patients}
The recruitment criteria for patients required them to be above the age of 70 (i.e., in the high-risk group for POD) and have undergone surgery in the preceding year. 
Recruitment was realized via the collaborating university clinic, which had pre-established contact with a suitable patient population. If they agreed, a first phone call with the participants was used to explain the study's research objectives, gather informed consent, and schedule an appointment for the telephone interview. Participants did not receive monetary compensation for participation.
Out of 78 contracted patients, 15 respondents agreed to partake in interviews conducted between November 2024 and February 2025.
Three patient participants dropped out of the study after the first phone call due to declining health conditions.

\subsection{Interviews}
\label{sec:method_interviews}

\subsubsection{Semi-Structured Background Interviews}
To gain an encompassing understanding of the patient journey of older adults undergoing surgery (i.e., in the high-risk group for post-operative delirium) along with the present implementation of POD-prevention measures within the application context (see \autoref{sec:context}), a series of semi-structured interviews with relevant stakeholder groups (see \autoref{tab:all_participants}) was conducted. 
The collected material is analyzed through a summarizing qualitative analysis procedure (for details see \citet{mayring_qualitative_2014}, p. 65 - 78). The results are used to construct a typical patient journey considering the possibilities for POD prevention and treatment measures, used as a guiding structure for the subsequent patient interviews. 
The interview guide for the semi-structured interview is designed to ensure the conversation remains focused on the topics of interest while allowing interviewees the flexibility to introduce new themes and expand on their responses. The appendix contains the interview guides for medical professionals (see \autoref{sec:questions_mp}) and patient representatives (see \autoref{sec:interview_guide_patients}).
The interviews were conducted as video calls using the application \textit{WebEx}. A total of eleven interviews were conducted between February 2024 and November 2024 and lasted, on average, 38.5 minutes. 
All interviews were audio recorded and transcribed verbatim.

\subsubsection{Patient Value Journey Interviews}
The objective of the patient value journey interviews is to review patients' experiences of undergoing surgery and elicit patient values relevant to their engagement and understanding of POD.
When working with vulnerable populations, research work across HCI and CSCW most commonly applies a participatory design approach ~\cite{anuyah_characterizing_2023}. While such approaches can promote deeper connections with vulnerable groups by positioning stakeholders as active participants in the design process, they also entail notable drawbacks as they may impose too great a burden on some participant groups~\cite{anuyah_characterizing_2023, brett_systematic_2014} or be overly technosolutionist~\cite{so_they_2024}.
This is particularly relevant for a vulnerable target group such as elderly patients, whose medical history and personal circumstances may make the typical effort of participating in such a design project too demanding. 
This sentiment was echoed in feedback encountered in early communications with older patients through our recruitment outreach, which indicated that our target group could not reasonably be expected to participate in participatory design activities as a the majority of prospective patients participants would consider the required efforts for such approaches (e.g., reading printed material, physical presence, or extended periods of engagement in workshop activities) excessively demanding and burdensome. 
\citet{grunloh_and_2024} already reported from first applications of the \textit{patient value journey mapping} (see discussion in \autoref{sec:vsd}) approach with patients that this method had a high cognitive and physical burden for the participants, proposing a condensed version thereof.
Based on these preceding methodological reflections and the patient feedback, we decided to devise an adapted version of the \textit{individual patient value journey mapping} approach~\cite{grunloh_and_2024, bui_patient_2023} into a semi-structured patient interview format that can also be conducted as a telephone conversation.
A pretest was conducted with two members of the research group who previously underwent surgeries, leading to improvements in the wording of the patient questionnaire.
A detailed account of the interview procedure and the guiding material is available in \autoref{sec:interview_guide_patients}.

\begin{table}[ht]
\caption{Overview of patient respondents for the patient value journey interviews, along with a comparison concerning five characteristics of their patient journey. Specifically, (1) whether a POD occurred, (2) if they recall receiving POD information in the pre-operative briefing session, (3) if devices to access digital media were available, (4) if the patients engaged in self guided research concerning their surgery, and if (5) relatives were involved in their hospital stay.}
\centering
\begin{tabular}{lccccc}
\toprule
\textbf{ID} & \textbf{(1) POD} & \textbf{ (2) Briefing Recall } & \textbf{ (3) Digital Media} & \textbf{ (4) Research} & \textbf{ (5) Relatives} \\
\midrule
P01 &  &  & \Checkmark & & \Checkmark \\
P02 &  &  & \Checkmark & \Checkmark & \Checkmark  \\
P03 &  &  & \Checkmark & \Checkmark & \Checkmark  \\
P04 &  &  & \Checkmark & \Checkmark & \Checkmark \\
P05 & \Checkmark &  & \Checkmark & \Checkmark & \Checkmark \\
P06 &  & \Checkmark & \Checkmark & \Checkmark & \Checkmark \\
P07 &  &  &  & \Checkmark & \Checkmark\\
P08 &  & \Checkmark & \Checkmark & \Checkmark & \Checkmark \\
P09 &  &  & \Checkmark &  & \Checkmark \\
P10 &  & & \Checkmark &  & \Checkmark \\
P11 & \Checkmark &  &  &  & \Checkmark \\
P12 & \Checkmark &  & \Checkmark & \Checkmark  & \Checkmark \\
\bottomrule
\end{tabular}
\label{tab:patient_participants}
\end{table}

\subsection{Thematic Analysis}
\label{sec:analysis}
This analysis of the \textit{patient value journey}~\cite{bui_patient_2023,grunloh_and_2024} aims to identify what values are of relevance for patients across their patient journey, consequently comparing them to the way they are embedded in the existing implementation in the research context and deriving design considerations for supporting patient empowerment in engaging in POD-preventive measures.

The analysis of the transcribed patient interviews is achieved through an inductive coding approach, specifically following Braun and Clarke's~\cite{braun_reflecting_2019, braun_can_2021, braun_critical_2024} process of six recursive phases for reflexive thematic analysis (TA).
This approach follows \citet{bui_patient_2023}'s inductive approach to analyzing patient value interviews, arguing that this allows for revealing implicit values underlying patients’ statements, as explicit verbalization of abstract values may be difficult for participants.
The TA process is as follows:
After initial \textbf{(1) familiarization} with the interview transcripts, all instances relevant to the RQ  are \textbf{(2) coded}. In this section, labels are defined and assigned with the purpose of encapsulating and evoking salient data features pertinent to addressing the research question. 
This process entails iteratively coding the entire dataset until a first consolidation of all codes and pertinent data extracts is achieved through the \textbf{(3) generation of initial themes}. 
These initial themes represent the main topics present in patients' retelling and reflection on their experiences concerning POD-preventive measures throughout their patient journey. 
Next, through \textbf{ (4) reviewing and developing candidate themes} against the dataset, deriving themes that present patterns of shared meaning underpinned by a central concept or idea, uniting implicit or latent meaning~\cite{braun_can_2021}.
These candidate themes are \textbf{(5) refined, defined, and named} by developing a detailed analysis of each theme, delineating its scope and focus. At this phase, each theme represents a patient barrier or value in the context of POD prevention.
Human values are thereby not understood to be absolute but rather an approximation of a concept. For this, they are not necessarily defined in a single word~\cite{davis_value_2015}.
In the final step, the values are \textbf{(6) written up} into an analytic narrative. This includes a description of barriers experienced by patients, highlighting values expressed throughout the patient journey, contextualizing them with the information derived from the background interviews, and categorizing them in terms of design requirements needed to embed these values in future digital applications for supporting patient engagement in POD prevention~\cite{hook_strong_2012}.
Following \citet{davis_value_2015}'s recommendation for reporting findings from VSD research, we choose to enrich our reporting with direct quotations from the participants. This allows readers to engage directly with stakeholders' choice of words and reduces the risk of researchers unintentionally reporting their own values and thoughts through summarizing and paraphrasing.

\section{Findings}
\label{sec:findings}
In the following, we first present the patient journey phases derived from the background interviews, before outlining the barriers and values identified in the thematic analysis of the patient value journey interviews.
\subsection{Patient Journey Phases}
Based on the interviews with medical professionals, we identified the following three distinct phases of a patient journey relevant to a patient's engagement with POD. These phases differ in the activities and degree to which the patient can actively engage in POD preventive measures, the information relevant, and the involvement of other stakeholders.
The three phases are: (1) Information and preparation, (2) Hospitalization and surgery, and (3)Post-hospitalization. Their differences concerning POD and related measures are summarized in the following:
\begin{enumerate}
\item \textbf{Information and Preparation.} Before being admitted to a hospital for surgery, the patient receives information about the possible occurrence of postoperative delirium, along with instructions, and can engage in preparative measures to reduce the risk of a POD occurrence.
\item \textbf{Hospitalization and Surgery.} Patients are admitted to the hospital and undergo surgery. POD may occur. Monitoring and engagement with the medical personnel are the highest, yet active engagement from the patient side is still possible. 
\item \textbf{Post-Hospital Stay:} Following hospitalization, the responsibility for undergoing rehabilitative treatment lies with the individual patient. 
\end{enumerate}

\subsection{Barriers across the Patient Journey}
\label{sec:barriers}
The following section will first delineate the barriers patients describe having experienced concerning POD and related treatment and prevention measures throughout their patient journey. 
%

\begin{figure*}[h]
 \includegraphics[width=\textwidth]{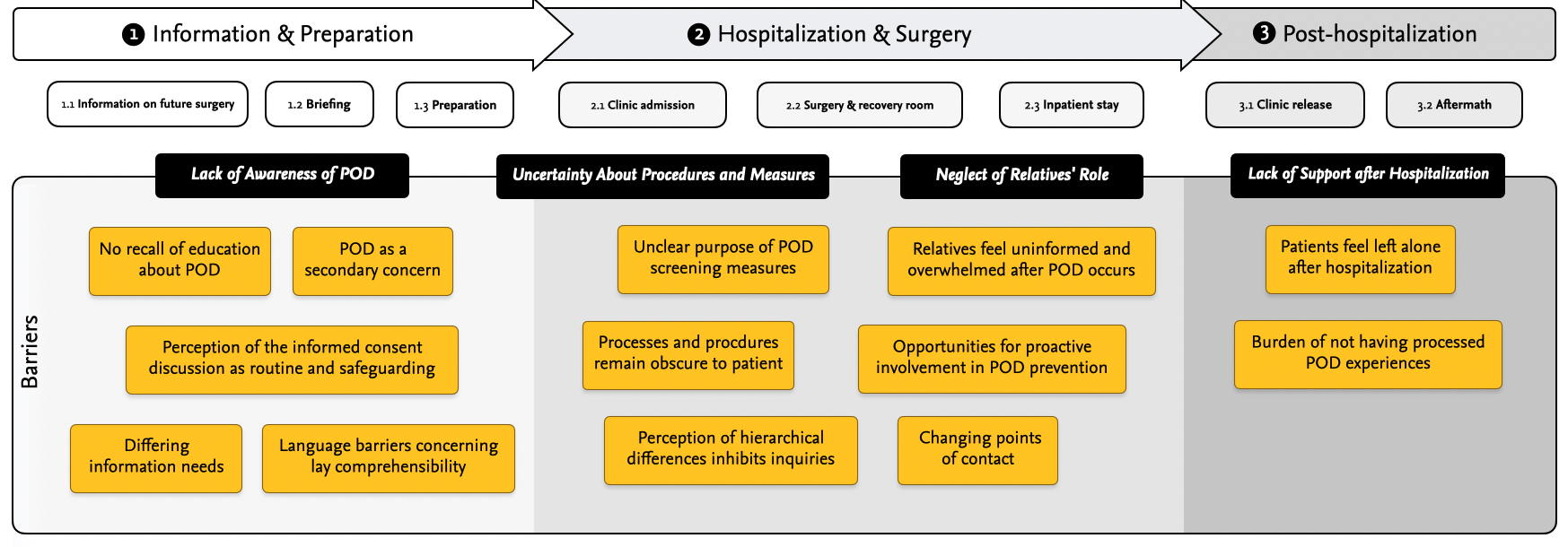}
  \caption{Overview of the identified barriers patients from the high-risk group for POD experience when undergoing surgery, including selected codes pertaining to these barriers.}
  \label{fig:barriers}
  \Description[]{The patients experience four main barriers across the patient journey. First, a lack of awareness of POD in the pre-hospitalization phase. Second, uncertainty about procedures and measures concerning POD throughout their hospitalization. Third, a lack of involvement of their relatives, and fourth, a lack of support after hospitalization in dealing with the consequences of a POD occurrence.}
\end{figure*}

\subsubsection{\textbf{Lack of Awareness of POD}} 
The patient interviews indicated a lack of awareness of POD, leading to a perception of uncertainty concerning POD, as they do not feel appropriately informed about associated risks, preventive strategies, and possibilities of active engagement after a POD occurrence.
Of the twelve interviewees, only two recall receiving information about POD throughout their patient journey (see \autoref{tab:patient_participants}).
Our interviews with health professionals identified the pre-surgery consultation between doctor and patient as the primary source of information for patients about POD. Patients, on the other hand, describe low levels of engagement in these briefing sessions, which they often consider to be merely formal, designed to provide legal safeguarding for the medical institution.
One patient describes this as follows: 
\begin{description}
   \item \emph{``So, about that preparatory meeting [...] I had a feeling it was a safeguarding meeting. There's a document with risks in it, and it's got to be filled out and signed off. To me, it feels like it's just an alibi function, and it's only there to protect the organization.''} (P04)
\end{description} 
Another patient who experienced POD shares this perception:
\begin{description}
   \item \emph{``What's the best way to communicate this [i.e., information concerning POD] at a time when you might have other worries? I think that's a bit difficult. Basically, the conversation is only there to provide reassurance. It's not intended to give the patient any decision-making possibilities or anything like that; it's just to cover the clinic, the surgeon.''} (P05)
\end{description} 
Given the amount of information (and its importance) conveyed in the informed consent discussion, the risk of POD is of secondary importance to patients.
Furthermore, the circumstances in which a briefing takes place may hinder patients from comprehending the information provided. A case in point is P12, who was rushed to surgery following an accident and describes not recalling briefing information due to the pain, \textit{``Yes, well, I was in so much pain that any information I was given disappeared.'' (P12)}
Additionally, varying informational needs and language barriers further hinder comprehension, making it difficult for patients to grasp the significance of POD. 
In this regard, patients report that the language used throughout the hospital stay remains inaccessible to them.
The overarching difficulty of obtaining information with treating physicians extends into the hospitalization phase, where one patient described the difficulty in engaging in conversations concerning their individual situations as follows:
\begin{description}
   \item \emph{``There's a new doctor every day. They're all under a lot of pressure, as you can imagine. It's almost become normal, unfortunately. But if you want to ask something, you must run after the doctor!''} (P01)
\end{description}
Furthermore, the perception of hierarchical differences toward healthcare professionals may keep patients from making use of this information offering, a factor that is aggravated by temporal stressors within the meeting situations and the use of medical language, furthering a perception of patients as laypeople (e.g., \textit{"[...] and you don't understand the medical terms the doctors use." (P11)})
These difficulties in gaining a larger understanding of processes like POD screening reinforce the conceptualization of patients as passive recipients of treatment measures, inhibiting active engagement.
\begin{description}\item \emph{``Since you [i.e., the patient] are such a layman in the field, you accept what you are told at face value, and that was also good for me.''} (P11)\end{description} 
These barriers to communicating POD-prevention information result in a lack of knowledge of POD or a disregard of it as a secondary, minor issue rather than a serious postoperative complication, commonly overshadowed by health worries.
Consequently, patients remain unclear about the possibility of improving their health through proactive engagement (i.e., in POD prevention activities).

\subsubsection{\textbf{Uncertainty About Procedures and Measures}} 
The patient interviews revealed that patients struggled with uncertainty about the purpose of POD measures and their role within prevention.
For example, P09 describes the patient perception of POD prevention measures as follows:
\begin{description}\item \emph{``I assumed they wanted to know if I was in my right mind. But it wasn't explained.''} (P09)\end{description} 
Specifically, patients lacked knowledge about the possibility of proactive participation in preventive activities through measures such as exercise or bringing auxiliary tools (e.g., hearing aids, glasses, or smart devices)  to the clinic.
This uncertainty directly results from the lack of knowledge on POD, discussed above.
Furthermore, a lack of clarity on the procedures experienced can also lead to uncertainty or worry on the patients' side. 
Changing points of contact and inconsistent communication further obscure patients' insights, leaving them unsure of whom to approach with concerns. 
The interviews reveal how patients perceive interactions with healthcare providers as fuzzy, as time constraints and fragmented communication chains make it difficult for them to gain a clear and encompassing understanding of the experienced care procedures. 
While patients largely recognize the effort and workload pertaining to the physician position, the rotating physician duty shift system introduces fragmentation in communication processes, resulting in patients facing difficulties in obtaining information. One patient describes this as follows:
\begin{description}\item \emph{``A different doctor came to see me every day. You can't tell them apart. [...] As I said, they all made an effort. [...] And the doctors are all under stress, of course. It's practically normal, unfortunately, but if you want to ask something, you have to run after the doctor, more or less.''} (P01)\end{description}
P02 reports a similar experience:
\begin{description}\item \emph{``I actually expected that there would be more doctors visiting the patients and I didn't recognize them either, there were different doctors there but they hadn't introduced themselves and they didn't have a tag on their chest. [...] I think it was a general shortcoming that neither the nurses nor the doctors had name badges, so you didn't know who you were dealing with. A different doctor came to see me every day. You can't tell them apart. [...] As I said, they all made an effort. [...] And the doctors are all under stress, of course. It's practically normal, unfortunately, but if you want to ask something, you have to run after the doctor, more or less.''} (P02)\end{description}
If this multitude of contact points with healthcare staff reveals perceived inconsistencies in the information provided, patients can experience uncertainty or anxiety concerning their treatment and health status. P04 describes the following:
\begin{description}\item \emph{``You get the feeling that they don't talk to each other. So, they don't have a good handover? There's obviously room for improvement in communication, and that's worrying because it can't just be the removal of a cannula; it could be something more serious. [...] So I found the communication during the stay to be a stressful factor when you realize that they don't agree among each other. So when they discuss it in front of you or discuss it in a way that you can hear it, it makes me feel anxious.''} (P04)\end{description}
If time constraints due to stress are signaled to patients, they may refrain from future inquiries. 
Beyond, the perception of hierarchical structures within the exchange with healthcare professionals further inhibits patients from seeking clarification, as they may hesitate to ask questions or challenge medical authority. This uncertainty can lead to missed opportunities for both early intervention and long-term preventive strategies. 
Finally, after hospitalization, this uncertainty can expand to rehabilitation measures. One patient describes this experience as follows:
\begin{description}\item \emph{``But at home I was already lacking confidence in how I was allowed to move, what I could do and what I shouldn't do, and you are not very self-sufficient. And you have to do everything yourself. It's so difficult to get dressed on your own that at that point, yes, I was downright lost.''} (P10)\end{description}

\subsubsection{\textbf{Lack of Support after Hospitalization}}
Some patients report feeling left alone after being discharged from the hospital, unsure about what next steps are required of them to take charge of their rehabilitation. One patient describes feeling overwhelmed by the information provided, as they missed instructions on how to use it.
\begin{description}\item \emph{``I got a big file from the physiotherapist with the letter of release [...]. It was handed over without comment, I just looked through it afterwards and was amazed at what they had supposedly done to me, but I didn't know anything about it [...]. I was just handed a sheet of paper with instructions on how to stand and all that. I mean, I knew my way around, but for someone who doesn't know anything, that's a lot. When you're just handed the paper like that, without any comment, it makes me wonder what the deal is.''} (P08)\end{description}
In particular, those patients who have experienced a POD may feel a sense of abandonment after such an occurrence, lacking the necessary guidance to cope with their experiences.
The absence of structured post-discharge support further exacerbates these challenges, leaving patients to grapple with the consequences of POD in isolation. 
Combined with a lack of knowledge about POD, patients can be left feeling helpless while having to deal with the psychological consequences of experiencing POD.
In this regard, one patient who experienced a particularly grave POD expressed a desire for healthcare professionals to acknowledge and address the experience and to provide guidance toward accessing psychological support to process the emotional and cognitive impact.
\begin{description}\item \emph{``Nobody talked to me about my condition and about the delirium or anything like that. I didn't even hear the term; I only learned it when a home care counselor used it. 'Oh, post-operative delirium', she said. I looked it up first; I'd been home from the hospital for a long time by then. I didn't know what it was. [...] And I am a person who does everything with my mind. I only had feelings for this whole experience and only feelings of horror. It would have been easier for me to work through it with my mind. But since I didn't have the concept, I couldn't do any good research.''} (P05)\end{description}
Here, the combination of a lack of knowledge and a lack of support after the occurrence of a POD inhibits the patient from processing the experience. Providing information on POD would support the patients in processing their experiences and enable those who seek to proactively address these through counseling to navigate toward the right specialists. 
%

\subsubsection{\textbf{Neglect of Relatives' Role}}
Following the occurrence of POD, patients' relatives find themselves in a mediating role between the patients and the clinic staff. This unanticipated circumstance often leaves them in emotional distress, feeling overwhelmed and lacking the information and guidance to best support the patient.
In addition, witnessing a loved one with POD can be a challenging experience for family members. The subjects encountered and expectations raised by interacting with a patient experiencing POD may be unexpected and intense, and include a dilemma in navigating and handling questions of guilt in dealing with the effects of the distorted experience of reality.
One patient who experienced POD describes how such experiences resulted in difficulties in the relationship with a close relative: 
\begin{description}\item \emph{``I ripped the tubes out because I wanted to escape. I wanted the police to pick me up because they weren't allowed to hold me against my will. And then they spoke to me and said [name redacted] is here, that's my granddaughter, and I said, yes, she wants to murder me. [...] So I'm still shocked that my granddaughter took offense at that. When I returned to reality, she took the first, best moment to tell me that I had reproached her and how bad it was, and I didn't know what to say. [...] I couldn't apologize for it either. I couldn't help it if I was in a world where everyone wanted to murder me. ''} (P05)\end{description} 
Relatives frequently are engaged in supporting patients, thereby compensating for the gaps created by uncertainty about processes. The following patient quote exemplifies such a situation:
\begin{description}\item \emph{``
I was discharged at midday on Saturday and I quickly called my wife and she picked me up. But if I hadn't been able to do that, I wouldn't have known how to get home, because there was no information that an ambulance would take me or that it could be a cab ride, so the information wasn't just thin, it wasn't there at all.''} (P02)\end{description}

In summary, we find that surgical patients are largely unaware of POD, leading to uncertainty around related procedures and limiting opportunities for patient engagement. As a result, patients often remain in a passive role. When POD does occur, both patients and their relatives struggle to cope with its consequences."

\subsection{Patient Values}
\label{values}
\begin{figure*}[h]
 \includegraphics[width=\textwidth]{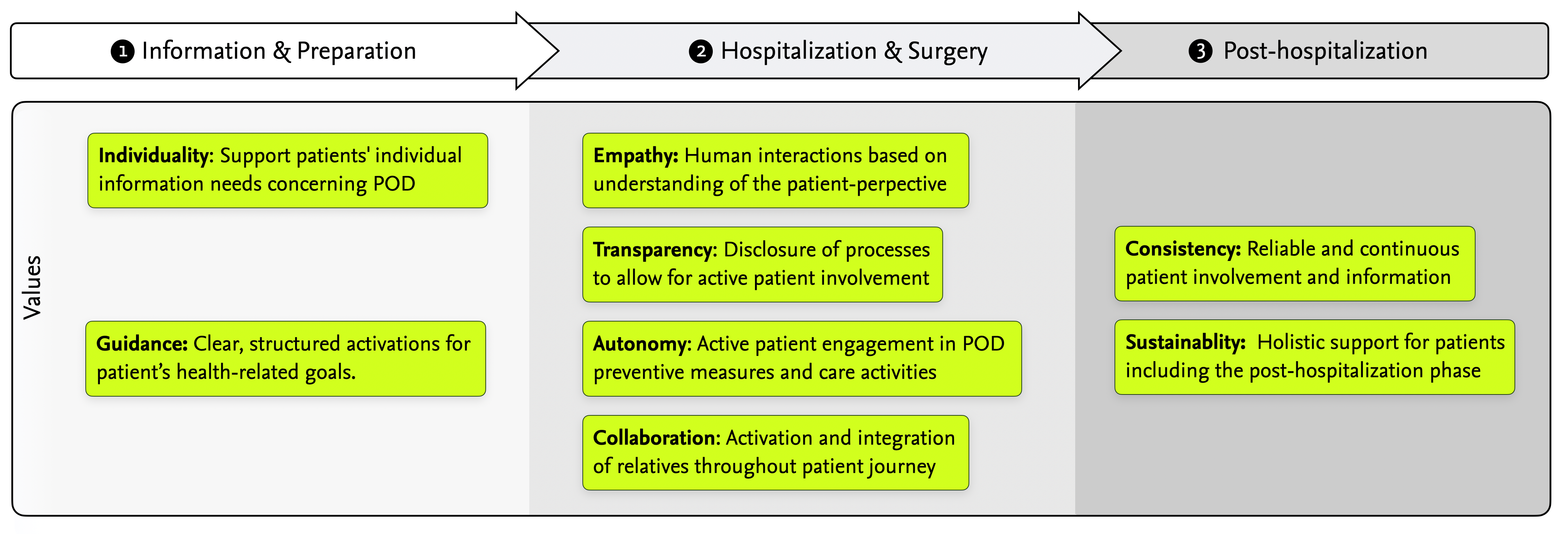}
  \caption{Overview of the identified patient values across the patient value journey.}
  \label{fig:values}
  \Description[]{The thematic analysis identified eight patient values across the patient journey, namely: Individuality, empathy, autonomy, sustainability, transparency, consistency, guidance and collaboration.}
\end{figure*}

The following section describes the values identified in the thematic analysis (see ~\autoref{sec:analysis}) of the patient value journey interviews.
This results section focuses on patients' expressions of values derived from their recounting and reflecting on their surgery experience.
As the values derive from the value patient journey interviews, in the value overview graphic (see \autoref{fig:values}) they are situated within the particular phase in which they have been most prominently discussed. Yet, as human values aren't clearly delineated concepts~\cite{friedman_value_2019}, their relevance is not strictly constricted to phase allocation in the overview graphic (see \autoref{fig:values}).
Design requirements are derived from integrating patient barriers and values in the subsequent discussion section.

\subsubsection{\textbf{Individuality}}
The value \textit{individuality} captures the need for communication modes and materials to be supportive with respect to an individual patient's differing information needs concerning POD. 
In the context of POD, this entails tailoring communication to accommodate differing levels of health literacy, personal relevance, and readiness for information across patients, as we found that patients differed in their motivation to seek information about their surgery. 
Within our sample, patients expressed two contrasting positions: They either engaged in a proactive inquiry of information or avoided the confrontation with additional information about the surgery.
Those patients who engaged in pro-active inquiry wanted to discover details about the upcoming operation and conducted independent research online. In this regard, we found that all the patients interviewed had access to online information except for two participants (P07, P11).
Patients in the avoidant information-seeking stance did not seek additional information before the surgery. The motivation behind this avoidance differs, as either anxiety or a stoic position toward their health may inhibit active information seeking.
Patients who showed a stoic stance are content that the prospective surgery addresses a health issue without strong emotional reactions. For example, P01 falls into this category:
\begin{description}\item \emph{``The doctor explained it to me, and looking back, I realized I hadn't done my own research before the surgery either. There's no point in googling it or anything else, it doesn't make any sense, at least not to me.[...] I take it as fate and I'm not going to do any more research. I'm not going to drive myself crazy.[...] I just let it happen, and I wasn't afraid. I thought, "Whatever comes, comes.''} (P01)\end{description}
 %
%

\subsubsection{\textbf{Empathy}}
The value \textit{empathy} describes patients' appreciation for communication (with healthcare personnel) that is rooted in an understanding of their situation on a human level.
%
%
Especially when POD occurs, the importance of human affection is highlighted by affected patients: 
\begin{description}\item \emph{``There was a nurse there. She was incredibly kind and she noticed that I was still not feeling well, so she held my hand. [...] I was still very confused and she helped me, let's say, to get back to normal life. I found the human touch very positive. You wouldn't believe what happens when someone holds your hand, or when someone is there and smiles and talks to you.''} (P12)\end{description}
Patients further highlighted how empathetic and calm interactions were perceived particularly positively, facilitating patient inquiry and consequently disarming the inhibiting effect of power differentials on patients' informedness about their health status and treatment options. P04 described the positive perception of eye-level conversations as follows:
\begin{description}\item \emph{``The best conversations were the ones where you were really open, I don't want to say at eye level, I'm not a doctor, but I can definitely understand what's going on and I can also ask questions and then get an answer on the same level.''} (P04)\end{description}

\subsubsection{\textbf{Autonomy}}
The concept of \textit{autonomy} signifies patients' aspiration for proactive involvement in decision-making and behaviors that can positively impact their health status. In the context of POD, this entails the possibility of meaningfully engaging in preventive measures aimed at reducing the risk of POD, facilitated through access to relevant information, resources, and opportunities for self-directed action throughout the patient journey.
P11 reports how, in their experience, concurrent structures required patients to relinquish their autonomy partly: 
\begin{description}\item \emph{``There is already a dissonance that makes people [i.e., patients] realize that there are some structures that I now need to commit to. In the best-case scenario, you have trust in the person you're talking to, and then that's just the way it is. But it's a bit like giving up your own autonomy.''} (P11)\end{description}

\subsubsection{\textbf{Sustainability}}
\textit{Sustainability}, as a patient value, describes the expressed need for holistic support of their active participation as the patient, not only throughout all phases of their patient journey, but also extending beyond the immediate period of hospitalization. Here, patients wish to receive instructions on what actions they need to take to achieve the best health outcomes, yet commonly lack knowledge thereof (see \autoref{fig:barriers})
\begin{description}\item \emph{``I personally found the instructions from the social care service really useful. They suggested some stuff I might need here  [i.e., at home post-hospitalization], like stocking pullers and a clamp. I wouldn't have known that, for example. I have to say, that was really helpful.''} (P07)\end{description}

\subsubsection{\textbf{Transparency}}
The patient value \textit{transparency} describes the disclosure of the processes involved in the prevention or treatment of POD delirium to the affected patients. This knowledge allows patients to become active stakeholders in their own recovery process. This stands in contrast to the passive patient role, where patients are the subjects of care measures that remain undisclosed and inaccessible to them. 
The value \textit{transparency} was primarily discussed in terms of uncertainty about procedures and measures. Hence, patient quotes for this value are found in the preceding barriers section (see \autoref{sec:barriers}).

\subsubsection{\textbf{Consistency}}
The value \textit{consistency} reflects patients’ expressed need for reliable and continuous support that is available in a similar degree if required throughout all stages of their experience of undergoing surgery. It emphasizes the importance of coherent and continuous patient involvement and information provision across all phases of the patient journey. 

\subsubsection{\textbf{Guidance}}
The value \textit{guidance} refers to patients’ appreciation for processes that provide clear and structured instructions on how to act in alignment with their health-related goals. 
For POD, this includes, for example, preparatory measures the patient can take to help reduce the risk of POD before surgery or instructions on how to best engage in rehabilitation measures.
P08 describes how they would have wanted better guidance on how to engage in rehabilitation after being discharged from the hospital:
\begin{description}\item \emph{``I would've liked a more personal approach. I would've liked to have been told how to get up and stuff. If you don't know that, you're left in the dark, and in that respect, I didn't think it was okay.''} (P08)\end{description}

\subsubsection{\textbf{Collaboration}}
The value \textit{collaboration} encompasses the meaningful involvement of relatives as important stakeholders in the patient’s care experience. This value reflects patients’ recognition of the relational and communicative support provided by relatives.
In this regard, all interviewed patients reported that a relative was involved in their hospitalization, taking crucial roles in communicating with both the patient and healthcare providers and supporting processes before, during, and after the hospitalization. 
This involvement takes different forms and sustains different functions. P04 describes how different family members provided support in different ways:
\begin{description}\item \emph{``My wife was the most involved. So she regularly inquired, kept an eye on things, and would ask whether I had already asked certain questions or not. She was very involved. My sons also came to visit and then also asked me how I was. They also brought a bit of everyday life back to my situation. So I was very well supported there, also on an emotional level. The bad thing is, you also tell them how you're doing. That alone reduces certain negative feelings. So you say, yes, I'm realizing what my health situation is, what I am [i.e., not healthy]. And you just have to be careful not to burden your loved ones too much with it.''} (P04)\end{description}
Once a POD occurs, the relatives' role becomes even more central. Then, relatives' involvement can play a crucial role in understanding and addressing changes in a patient's behavior. For example, P01 describes how her daughter detected the onset of a POD.
\begin{description}\item \emph{``Yes, well, you know, I didn't notice it [i.e., POD] at all. But my daughter said that she did notice that I was a bit depressed.''} (P10)\end{description}
Furthermore, patients also expressed comfort in knowing that relatives were actively engaged by the involved healthcare staff:
\begin{description}\item \emph{``And what I found very comforting was that as soon as the operation was carried out, [the surgeon] quickly informed my wife that everything had gone well.''} (P09)\end{description}

\section{Discussion}
\label{sec:discussion}
To approach our guiding research question (\textit{What are patients' values and barriers concerning POD and related prevention measures when undergoing surgery?}), we first conducted semi-structured interviews with clinical stakeholders and patient representatives to gain an comprehensive understanding of the typical patient journey for older adults in the high risk population for POD. 
Based on this conceptualization, we conducted patient value journey interviews with twelve elderly surgical patients to develop a nuanced understanding of older adults’ experiences with POD and its prevention.
Reflexive thematic analysis revealed barriers and values from the patient perspective, offering insight into how current POD-prevention practices with patient priorities.
Addressing our second research question (\emph{What are the design requirements for a digital health application that supports patients' values towards overcoming barriers towards engagement in POD-preventive measures?}) we derive design requirements in the form of \textit{`intermediate level design knowledge'}~\cite{hook_strong_2012}. 
This is achieved by operationalizing the identified patient values (see \autoref{sec:values}) against the existing barriers (see \autoref{sec:barriers}) into \textit{strong concepts}~\cite{hook_strong_2012} for the future design of digital health applications supporting patients in POD (discussed below, a schematic overview is provided in \autoref{fig:synthesis}).
Through this, we articulate design knowledge, grounded in specific application contexts, which seeks to be generative towards the future design of sociotechnical systems that aim to support patient agency and engagement in CSCW and the digital health domain ~\cite{hook_strong_2012, schmidt_permutations_2007, fitzpatrick_review_2013}.
The four identified design requirements are interconnected, as they are most supportive toward patient support in an integrated fashion. 

\begin{figure*}[h]
 \includegraphics[width=\textwidth]{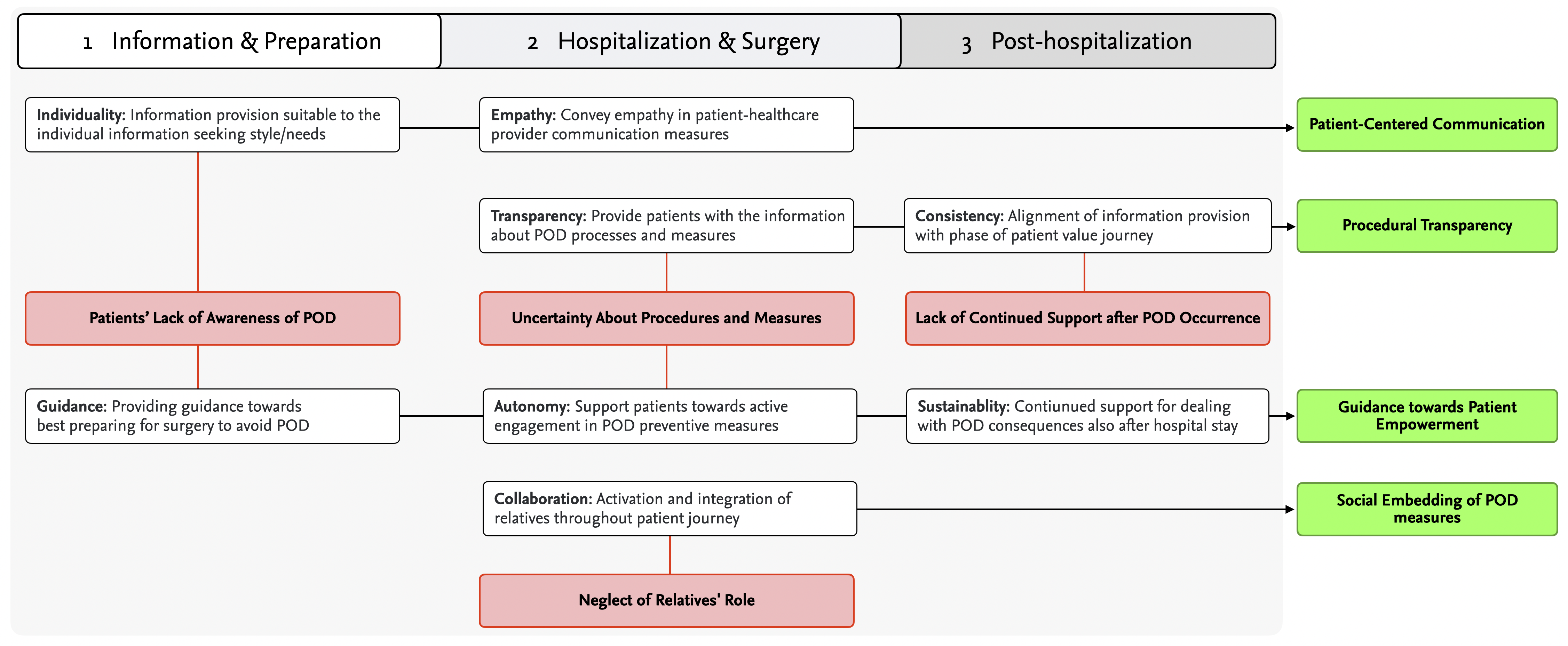}
  \caption{Overview of the operationalization of design requirements from patient values and barriers across the patient value journey.}
  \label{fig:synthesi}
  \Description[]{The figure illustrates key patient values, namely individuality, empathy, transparency, consistency, guidance, autonomy, sustainability, and collaboration, alongside identified patient barriers, including a lack of awareness about POD (patient outcomes data), uncertainty about procedures and measures, insufficient continued support after POD occurrences, and the neglect of the role of relatives. The design requirements outline how these values can be operationalized to address these barriers. They include patient-centered communication, procedural transparency, guidance toward patient empowerment, and social integration of POD measures.}
\end{figure*}

\subsection{Patient-Centered Communication}
\label{sec:requirement_communication}
To address the lack of awareness of POD from the patients' side, a consideration of the values \textit{individuality} and \textit{empathy} is needed. Based on these two values, we propose the design requirement of \textit{Patient-centered communication}, which is subsequently discussed in terms of designing digital health applications.
To realize \textit{`empathy'} in patient communication measures, conceptualization of patient communication material (e.g., an information portal concerning education on POPD) must be rooted in understanding the patient experience. This requires information to be presented in a fashion that is empathetic to a patient's sociocultural context (i.e., in terms of language) and individual situation (i.e., level of pre-operative anxiety).
To address \textit{`individuality'}, that is, the diverging information needs of individual patients, information provision strategies must seek to provide information that is relevant to a patient in a format that is respectful of their information needs.
Hence, we derive the following design requirement:
\begin{leftbar}
\begin{coloredleftbar}{black!10}
        \textbf{Patient-Centered Communication:} \par Provide information in a \textit{patient-centered content format} relevant to individual patients’ \par information needs and the patient experience to create awareness on  POD-related matters.
\end{coloredleftbar}
\end{leftbar}

The aim of this design requirement is twofold: First, it ought to address the lack of knowledge concerning POD identified in the patient interviews by providing information in a most suitable format to the individual patient, and second, to support the patient-provider communication on POD.

\paragraph{Informing Individual Patients on POD}
While the differing information-seeking behaviors identified in our patient interviews (e.g., stoicism, avoidance, and proactive inquiry) may not be comprehensive, they demonstrate the individual differences in information needs within the target group.
This aligns with previous research~\cite{kassam_patient_2023, karimi_scribe_2021}, which suggests that patients have different information needs that necessitate tailored information provision.
\citet{xue_preoperative_2020} work provides further insight, finding the effect of individualized preoperative information approaches on POD incidences, as patients undergoing cardiac surgery have a lower incidence of POD if individuals received individualized education sessions, a critical care tour, as well as an information leaflet, compared to those receiving standard care. 
These findings demonstrate that for patients, access to relevant health information can serve to support decision-making for health and well-being as they shape how a medical situation is experienced~\cite{patel_information-seeking_2022}.
But beyond access, the information must be provided in a format that is effectively accessible to patients (i.e., in terms of length, language, accessibility), as otherwise it fails to be considered by patients.
Concerning the language used in educational material, \citet{park_patient_2017} underlines how the use of medical jargon withholds patients from fully understanding elements of their care process. 
%
%
This finding aligns with patients' reports that medical language presents a barrier to their understanding of POD in the present study. 
Working with a comparable patient group, \citet{hao_exploratory_2023} described how older adults reported difficulties comprehending technical terms related to their conditions, leading to confusion and the inability to engage in healthcare processes actively. The authors argue that simplifying technical language and providing more detailed explanations would enhance patients’ understanding of their treatment options, enabling them to participate more actively in their care. 
%
%
%
Furthermore, it must be considered, that patients' information needs can further change and develop along the patient journey: \citet{patel_information-seeking_2022} points out that as people are exposed to new information about their health (e.g., that a surgery is required), the explanatory model of their illness experience changes along with their information-seeking behavior.
Digital health applications, such as electronic health portals, offer a means to support patients in engaging with information at the level of detail required by each individual patient and in an accessible format~\cite{karimi_scribe_2021}. 
Work from HCI and the Health Informatics domain found that better information access via digital health technologies can lead to increased adherence to care and patient empowerment~\cite{park_patient_2017}

\paragraph{Supporting Patient-Provider Communication}
In our analysis of patient barriers, we found that patients do not perceive the content of the information sessions as relevant to their specific situation (but either as legal cover or as familiar information). 
\citet{karimi_scribe_2021} described how older patients may face barriers and accessibility challenges that limit their access to shared information. Information that is provided during patient-physician conversations may not be available to patients after such a session ends, as patients find it challenging to handle multiple paper leaflets to find the required information or recall tailored information that has only been discussed verbally~\cite{karimi_scribe_2021}.
This aligns with the barriers identified in the patient interviews (see \autoref{sec:barriers}). A failure to provide a suitable information environment in the patient briefings results in a lack of patient awareness concerning POD and resulting uncertainty about related procedures and measures. Power differentials inhibit patient inquiry and make medical language inaccessible. 
In this regard, adopting a \textit{patient-centered communication} can bridge these differentials by supporting patients in their communication with healthcare providers and providing \textit{empathetic} support, rooted within an understanding of the patient perspective.
%
%
\citet{berry_getting_2017} reported that patients who are unable to establish a shared understanding with their healthcare providers regarding their health care priorities may experience poorer health outcomes and quality of life.
Further, the authors emphasize that while providers often aim to understand what matters to patients, they frequently apply this understanding to advance their own priorities, potentially overshadowing the patients' values. As a result, patient-provider communication about patient values remains limited~\cite{berry_getting_2017}. 
This may cause patients to perceive communication opportunities with providers as merely legal and procedural routines, rather than as chances to position themselves and their behavior optimally for proactive engagement in an upcoming care procedure (e.g., surgery).
Digital health applications have been proposed as a means to improve communication (and subsequently health outcomes) between healthcare providers and patients \cite{rajabiyazdi_differences_2017}, as they enable patients to access relevant information outside of one-on-one interactions, which have been shown to result in limited recall~\cite{karimi_scribe_2021}.

Taken together, a patient-centered communication approach tailored to individual information needs can help overcome communicative barriers by supporting personalized information delivery, thereby enhancing awareness of postoperative delirium (POD) and improving patients’ understanding of opportunities for active engagement in preventive measures.

\subsection{Guidance towards Patient Empowerment}
\label{sec:requirement_empower}
We found that the lack of awareness of POD on the patients' side leads them to assume a passive role in the further course of the patient journey (see \autoref{sec:barriers}).
A digital health application should \textit{guide} patients toward \textit{autonomous} engagement in POD prevention activities. This guidance needs to \textit{sustain} the entirety of the patient journey, including the phases before and after hospitalization. Through continued activations, patients should be empowered to take an active role in POD-related measures.
Hence, we operationalize the values \textit{guidance}, \textit{autonomy}, and \textit{sustainability} into the design requirement \textit{`Patient Empowerment'}, describing the provision of guidance to patients towards active engagement in their care processes. 
\begin{leftbar} 
\begin{coloredleftbar}{black!10}
        \textbf{Guidance towards Patient Empowerment:} \par \textit{Continued guidance} throughout the entirety of the patient journey, including post- \par hospitalization, providing  patients with \textit{consistent} activations supporting an active \par engagement in POD preventive measures.
\end{coloredleftbar} 
\end{leftbar}
This design requirement addresses two of the barriers that inhibit active patient-side engagement, namely, uncertainty about procedures and measures, and lack of awareness of POD.
These patient barriers match those observed by \citet{pfeifer_vardoulakis_using_2012} in their study on patients' perceptions of emergency room settings. Patients reported encountering unfamiliar procedures, delays in information provision, and complex terminology, which relegated them to a passive role and hindered their ability to make informed decisions or actively participate in their treatment. In this context, the authors identified a digital health application designed to provide guidance, thereby increasing patient empowerment.
\citet{park_patient_2017} further corroborated how encouraging patients to actively seek out more information can foster an engaged and empowered patient role.
From a medical perspective, such guidance towards active engagement in POD prevention measures further supports patients' autonomy by working towards safeguarding health, cognitive, functional, and mental functions~\cite{yurek_qualitatsvertrag_2023}.
\citet{swarbrick_evidence-based_2022}'s review of evidence-based strategies for POD prevention concluded that supporting individualized patient education is critical for patient-sided POD prevention, as it equips patients and their families with tailored information about the surgical patient journey, expected outcomes, and care environment. This approach was shown to significantly reduce POD incidence by promoting preparedness, reducing anxiety, and encouraging proactive engagement in preventive strategies. Hence, the authors argue for the implementation of multi-component POD-related interventions through the involvement of all peri-operative stakeholders, including patients and their families.
In accordance with these prior works, our design requirement posits patient empowerment through continued activation, rendering the possibilities of participating in their care apparent. Furthermore, the continuity of these activation measures must be supported throughout the entirety of the patient journey, spanning the time frame from the first consultation to the aftercare following hospitalization. 
%

\subsection{Procedural Transparency}
\label{sec:requirement_transparency}
To operationalize the values \textit{transparency} and \textit{consistency}, we posit the provision of information that is most relevant to a patient's current experience (i.e., the current phase of the patient journey), to address the uncertainty about POD procedures and measures, identified as a barrier to patient engagement therein.
Hence, we define the design requirement \textit{Procedural Transparency} as follows:
\begin{leftbar} 
\begin{coloredleftbar}{black!10}
        \textbf{Procedural Transparency:} \par \textit{Phase-specific tailoring} of the provided information to the current phase of the patient \par journey to enhance transparency and provide information at the moments it is needed.
\end{coloredleftbar}
\end{leftbar}

By fostering transparency on the care processes related to POD for patients through the design of a digital health application for patients, the patient's uncertainty should be addressed, to support them in taking an active role, and take an active role in their POD prevention and treatment.
Providing transparency further alleviates power differentials, affecting patients' agency during hospitalization.
According to \citet{dahl_facilitating_2020}, the sources for such asymmetrical power relationships between a healthcare professional and a patient can include differences in professional and institutional knowledge, experiential relevance, and understanding of institutional routines. 
In this regard, clinical routines may appear predictable and transparent to a healthcare worker, whereas they often remain unpredictable and opaque to a patient~\cite{dahl_facilitating_2020}. Therefore, patients and care providers are frequently unaware of such gaps in comprehension ~\cite{wilcox_designing_2010}. 
Hence, by offering procedural transparency, digital health applications have the potential to facilitate timely and appropriate support while enabling patients to identify ways to proactively engage in their care processes.

Beyond educational and informational needs, this further aligns with findings by Haldar et al.~\cite{haldar_beyond_2019}, who identified a range of open patient needs during hospitalization that remain insufficiently supported by current digital health tools (e.g., through patient portals), namely, the transition from home to hospital, adjusting schedules and receiving status updates, understanding and remembering care, asking questions and reporting problems, collaborating with healthcare professionals as well as preparing for discharge and at-home care after a hospital stay. 
All of these unmet needs pertain to procedures spanning the holistic patient journey, i.e., beyond the time frame of hospitalization. In this context, phase-specific tailoring of guidance ensures that patients receive relevant support at the appropriate time, rather than being overwhelmed with comprehensive information during a single briefing session held weeks before surgery (see \autoref{sec:barriers}).
Finally, circumstances along the patient journey can leave patients with situational impairments \cite{haldar_beyond_2019}, temporarily limiting their ability to engage actively. In this regard, the design of digital health technology must consider patients' potential for active engagement per phase~\cite{grunloh_and_2024}.

In summary, phase-specific procedural transparency equips patients with timely, relevant information, reducing uncertainty and power differentials and enhancing their ability to engage in their care process.

\subsection{Social Embedding of POD measures}
\label{sec:requirement_social}
In light of the lack of recognition, the role a patient's social environment plays in the POD patient journey (see \autoref{sec:barriers}), considering relatives' role as mediators and supporters in the patient-provider relationship is central to realizing the value \textit{collaboration}.
Hence, to operationalize the value \textit{collaboration} into a design requirement, we expand the need for context-specific information provision described above (see \autoref{sec:requirement_transparency}) through the requirement of role-specific information provision that recognizes the role and situation of relatives as active stakeholders.
Hence, a future digital health application should support the social embedding of POD prevention as follows:
\begin{leftbar} 
\begin{coloredleftbar}{black!10}
        \textbf{Social Embedding:} \par \textit{Role-specific tailoring} of digital information provision to the specific role of a person accessing \par it, including the explicit addressing of relatives in their mediating and supporting role before \par and after an occurrence of POD. 
\end{coloredleftbar}
\end{leftbar}
As relatives of POD patients find themselves thrown into unanticipated circumstances, they commonly feel overwhelmed and lack the information to best support the patient. 
This situation can prove strenuous for those affected: Relatives of POD patients were found to experience distress at high levels and increased workloads through the care of an affected family member~\cite{yurek_qualitatsvertrag_2023}, as well as symptoms of post-traumatic stress disorder, depression, and anxiety~\cite{swarbrick_evidence-based_2022}.
Yet, the involvement of a patient's social environment can be critical, as the involvement of relatives has previously been proposed as a non-pharmacological strategy within delirium management bundles. However, the addressing of relatives still lacks systematic implementation~\cite{lin_family_2024}.
A digital health application can alleviate this situation by rendering the information relatives require to adapt to these new situations accessible quickly and selectively, tailored to their specific role within treatment. 
By expanding the scope of guidance and information provision tailored to the role of the relative, they can be better prepared to fulfill their supportive function towards patients and healthcare providers, as well as better deal with these extraordinary circumstances affecting someone close to them:
To support patients, relatives can uphold familiar relationships and routines with the patient, which is an important POD treatment approach~\cite{swarbrick_evidence-based_2022} that has been further highlighted by the interviewed patients, as well as assist in organizing or advancing care procedures on the patient's behalf~\cite{karimi_scribe_2021}.
Concerning healthcare staff, recognizing and guiding the relative's role can aid communication with the patient and assist healthcare providers in identifying POD occurrences by reporting noticeable changes in patient behavior. 
Finally, improved preparation can help family members anticipate and manage potentially aggressive or extreme behavior induced by POD, which has been found to cause significant emotional distress, as well as symptoms of post-traumatic stress and depression among affected relatives~\cite{yurek_qualitatsvertrag_2023,swarbrick_evidence-based_2022}.

In summary, recognizing the role of relatives in a digital health application for POD prevention means providing them with guidance and information tailored to their role. The provision of such resources enables relatives to support patients at this critical time and to cope with the difficult circumstances themselves.


\section{Limitations}
Two main limitations should be considered for this study: 
First, due to the highly contextualized nature of complex healthcare settings and the difficulty of recruiting a vulnerable group within them, the patient sample size of fourteen remained limited \cite{fitzpatrick_review_2013}, which also limited generalizability. For this reason, we included patient representative interviews within the first phase to help reflect broader patient concerns and address the targeted level of generalizability by approaching the design requirements as strong concepts~\cite{hook_strong_2012}.
Second, as is characteristic of qualitative methodologies such as thematic analysis, the interpretation of data is shaped by the researchers’ perspectives. We recognize theme development in TA as an analytic and interpretative work on the part of the researcher~\cite{braun_reflecting_2019}, which may consequently be influenced by the coding researchers' positionality. For this reason, we provide a positionality statement (see \autoref{sec:positionality}).
These limitations underscore the importance of reflexivity and transparency in the analytical process, particularly in CSCW research that involves interactions with vulnerable populations.

\section{Conclusion}

This work aims to explore how design requirements can be derived for a patient-centered development of digital health technologies, focused on the application context POD. 
Our approach emphasizes researchers and designers' responsibility to center patient values in the design process in the digital health domain. Therein, the patient's role is often marginalized due to power imbalances between stakeholders, resulting in limited consideration of patient perspectives.
Values associated with patients’ individual experiences are frequently underrepresented in the design of digital technologies, as more influential stakeholders, such as medical practitioners, prioritize treatment-related values. Consequently, aspects of the patient’s experience, including their life philosophy and background, are often overlooked.
Patient-centered and value sensitive design approaches, as discussed in this paper, can serve to highlight these patient values and make them integral to future technology development, thereby facilitating patients' engagement in future digital health processes.
The present paper seeks to provide a blueprint for how such a patient-centered approach can be taken for basing the design of digital health applications on the values of a vulnerable stakeholder group (i.e., older patients about the age of 70).
In outlining our methodological approach, we provide a framework that can help researchers and designers to work towards the ethical alignment of digital health solutions and, through that, foster meaningful, human-centered interactions that support patient engagement and improve health outcomes.
Finally, by proposing design requirements in the form of \textit{strong concepts}, we provide actionable recommendations for the future design of patient-centered technology within the context of surgery-related hospitalization and POD.



\bibliographystyle{ACM-Reference-Format}
\bibliography{references}

\appendix

\section{Semi-Structured Interview Guides}

\subsection{Clinical Staff}
\label{sec:questions_mp}

 \textbf{Background of the Interviewee}
\begin{enumerate} 
    \item To begin with, could you briefly describe your professional background?
    \item In what way do you encounter postoperative delirium in your daily work?
\end{enumerate}

 \textbf{Questions about POD and POD Guidelines}
\begin{enumerate} 
    \item Could you describe a typical surgical process, focusing on the moments when postoperative delirium or preventive measures for it play a role?
    \item How can patients be supported with regard to POD?
    \item Where do you see the greatest challenges in the prevention or treatment of POD?
    \item How familiar are you with the POD guidelines? Please describe briefly.
\end{enumerate}

 \textbf{Questions about the Role of Patients}
\begin{enumerate} 
    \item What might a hypothetical digital system look like that would optimally support patients with regard to postoperative delirium?
    \item Is there anything else you think is important to mention concerning patients and a possible digital support system?
\end{enumerate}

\subsection{Patients Representatives}

 \textbf{Background of Interviewee}
\begin{enumerate}
    \item Please briefly describe your professional background.
    \item In what way have you encountered the topic of POD in your role?
\end{enumerate}

 \textbf{Questions about POD and POD Guidelines}
\begin{enumerate}
    \item Can you describe a typical surgery process from the patient's perspective?
    \item In your opinion, what are the challenges regarding POD from the patient's perspective?
    \begin{itemize}
        \item What information do patients need in your experience to help prevent POD?
        \item What information or support should be provided to patients by a clinic where surgery is performed?
        \item At what point should this information/support be provided for POD?
    \end{itemize}
    \item In which situations or contexts do you consider it important to support patients regarding POD risk after hospital discharge?
    \begin{itemize}
        \item What kind of information could help support discharged patients in the postoperative period?
    \end{itemize}
    \item What role do you think relatives play in communicating and ensuring adherence to recommendations for POD prevention?
\end{enumerate}

 \textbf{Questions about Planned Digital Decision Support System}
\begin{enumerate}
    \item How would you envision a digital support system that best helps patients follow POD prevention measures?
    \begin{itemize}
        \item What content should it include?
        \item What features could it additionally provide?
        \item What considerations should be made regarding the target group (70+, postoperative) in the design of a user interface?
    \end{itemize}
\end{enumerate}

\section{Interview: Patient Value Journeys}
\label{sec:interview_guide_patients}

\subsection{Overall Process}
\begin{enumerate}

    \item \textbf{Introduction} The participants are welcomed, get a quick reminder of the interview procedure (complete participant information has been sent and signed via post in the preceding weeks), and are provided with the possibility to address any open questions. 
    
    \item \textbf{Patient Journey Interview} 
    The first part of the interview serves to gain an understanding of a patient's experience of undergoing surgery, their emotional state throughout the phases of their patient journey, and the involvement of other stakeholders (e.g., relatives).
    Next, the patient participants are guided step-wise through the different phases of a typical patient journey for an older patient undergoing surgery (derived from the clinical staff interviews) and asked to describe their experience, their emotional state at the time and the involvement of other stakeholders per phase.
    For this, each phase is introduced through a short question prompting patients to recall their experience (e.g., for the initial phase:  \textit{"Firstly, let us discuss the moment you became aware that surgery was required. Please describe your experience of this"}).
    If a respondent takes the lead, the conversation follows their initiated flow. Subsequent questions then serve to fill in missing phases or content~\cite{grunloh_and_2024}. When patients do not initiate a flow of conversation, inquiries into the respondents' experiences per phase of the patient journey are prompted by the interviewer in chronological order.
    Finally, patients are asked if any critical events were missing from the discussed phases to allow for the introduction of additional steps in their patient journey. 
    
    \item \textbf{Value Elicitation} 
    Following this initial series of questions focused on the experience and emotions associated with a past surgery, each phase is revisited a second time, eliciting the values of importance to patients for each phase established in the first section.
    Seeking to cultivate the emergence and discovery of local and individual expressions of values~\cite{iversen_rekindling_2010}, an open-ended questions format is maintained, prompting patients to reflect on their experiences in terms of needs and preferences, initiating the conversation by asking: \textit{"During this phase, what form of support, assistance, or information would have been of importance to you?"}.
    The open question format is intended to reflect the comprehensive understanding of values within VSD as \textit{"what is important to people in their lives."}~\cite{friedman_survey_2017}, following the objective of eliciting a range of reflections and considerations from the patient for consideration in the thematic analysis (detailed in\autoref{sec:analysis}).
    Investigating emotions and incidents in such a fashion can help identify the underlying values~\cite{grunloh_and_2024}.
    If needed, the interviewing researcher asks follow-up questions, prompting patients to go into further detail.
    Furthermore, aspects pertaining to values might have already been mentioned in the preceding phase of the interview. In this case, the patients are prompted to go into detail and reflect on why and how a particular value was important for them during the respective moment.
\end{enumerate}


\end{document}